\documentclass[preprint,showpacs,preprintnumbers,amsmath,amssymb]{revtex4}
\usepackage{graphicx}
\usepackage{dcolumn}
\usepackage{bm}

\begin{document}

\title{Graphene-protected iron layer on Ni(111)}

\author{Yu. S. Dedkov$^{1,}$\footnote{Corresponding author. E-mail: dedkov@physik.phy.tu-dresden.de}, M. Fonin$^2$, U. R\"udiger$^2$, and C. Laubschat$^1$}
\affiliation{\mbox{$^1$Institut f\"ur Festk\"orperphysik, Technische Universit\"at Dresden, 01062 Dresden, Germany}\\
             \mbox{$^2$Fachbereich Physik, Universit\"at Konstanz, 78457 Konstanz, Germany}}

\date{\today}

\begin{abstract}
Here we report the photoemission studies of intercalation process of Fe underneath graphene layer on Ni(111). The process of intercalation was monitored via XPS of corresponding core levels and UPS of the graphene-derived $\pi$ states in the valence band. \textit{fcc}-Fe films with thickness of 2-5 monolayers at the interface between graphene and Ni(111) form epitaxial magnetic layer passivated from the reactive environment, like for example oxygen gas. 
\end{abstract}


\maketitle

Magnetic thin films with out-of-plane (or perpendicular) magnetic anisotropy play an important role in impetuous developed field of science as nanotechnology. Such systems can be used as a basis elements in perpendicular recording hard drives which became commercially available in 2005~\cite{Toshiba:2005}. Perpendicular recording is predicted to allow information densities of up to around 1\,Tbit/inch$^2$, a quadrupling of today's highest areal densities~\cite{Hitachi:2007}. The widely used materials with out-of-plane magnetic anisotropy are CoPt or FePt alloys. Recently, \textit{fcc} Fe thin films deposited on different substrates have attracted heightened interest for applications in novel 2D devices, like perpendicular recording, due to perpendicular anisotropy in these films. 

Recent calculations~\cite{Wang:1985,Krasko:1987,Moruzzi:1989} showed that, by stabilizing $3d$ transition metal elements in structural phases different from their natural ones, peculiar and generally enhanced magnetic properties could be obtained. In the case of Fe, such a behavior was shown for the low-temperature $\gamma$ phase (\textit{fcc}), which is naturally stable only at high temperatures for the bulk ($T>1183$\,K), but can exist at room temperature in thin epitaxial films grown on suitable \textit{fcc} substrates.

As for the study of magnetic properties, the necessity of a close lattice match between substrate and deposited material usually results in the choice of nonmagnetic \textit{fcc} templates like Cu~\cite{OBrien:1995,Wuttig:1995}. However, Fe films grown on different Cu surfaces have been found to show different growth modes and different ranges of thicknesses in which the ferromagnetic \textit{fcc} phase with out-of-plane anisotropy persists. Ni is another suitable material due to the small lattice mismatch of $+2$\%, referred to the room temperature lattice parameter of \textit{fcc} Fe, extrapolated from the high-temperature phase~\cite{Sander:1997,Johnston:1997,Arnold:1997,DAddato:2000,Gazzadi:2002}. It was shown~\cite{Sander:1997,Arnold:1997} that Fe layer on top Ni(111)/W(110) induces perpendicular to the film magnetic anisotropy for $0.5-3$\,ML thick iron films. At higher Fe coverages, an in-plane magnetization was found, which is proposed to be caused by the \textit{fcc} to \textit{bcc} transition in the Fe layer around 4\,ML thickness of iron film.

The aim of the present work is the preparation of a system with out-of-plane magnetic anisotropy which is passive to an aggressive environment. Here we demonstrate the possibility to prepare such a system via intercalation of thin Fe film underneath graphene layer formed on a Ni(111) substrate. Graphene layer behaves like a protection in this system conserving magnetic properties of the underlying epitaxial Fe film.

Investigations of the Fe intercalation were performed in the experimental setup for photoelectron spectroscopy consisting of two chambers described in detail elsewhere~\cite{Dedkov:2006,Dedkov:2007}. As a substrate the W(110) single crystal was used. Prior to preparation of the studied system the well established cleaning procedure of the W-substrate was applied [see Fig.\,1(a)]. A well ordered Ni(111) surface was prepared by thermal deposition of Ni films with a thickness of about 200\,\AA\ on to a clean W(110) substrate and subsequent annealing at 600\,K. The corresponding low-energy electron diffraction (LEED) pattern is shown in Fig.\,1(b). An ordered graphene overlayer was prepared via cracking of propene gas (C$_3$H$_6$) according to the recipe described in Ref.~\cite{Dedkov:2008a,Dedkov:2008b,Dedkov:2001}. The LEED spots of the graphene/Ni(111) system reveal a well-ordered $p(1\times1)$-overstructure as expected from the small lattice mismatch of only 1.3\% [Fig.\,1,(c)]. After the cracking procedure the Ni(111) surface is completely covered by the graphene film as was earlier demonstrated in Ref.~\cite{Dedkov:2008a,Dedkov:2008b}. Intercalation of Fe underneath graphene layer was performed via annealing of 2\,ML-thick predeposited Fe film at moderate temperature. Temperature was simultaneously monitored by optical pyrometer and W-Re thermocouple. Photoemission spectra monitoring the process of intercalation were recorded at 21.2, 40.8\,eV (He\,I$\alpha$, He\,II$\alpha$, UPS) and 1253.6, 1486.6\,eV (Mg\,K$\alpha$, Al\,K$\alpha$, XPS) photon energies using a hemispherical energy analyzer SPECS PHOIBOS 150. The energy resolution of the analyzer was set to 50 and 500\,meV for UPS and XPS, respectively.

Fig.\,1 shows LEED images of the successive steps of the preparation of the intercalation-like system on the basis Fe and graphene. The predeposited Fe film with the nominal thickness of 2\,ML on top of the graphene/Ni(111) system was annealed at $\approx600$\,K leading to Fe intercalation, which forms a commensurate $(1\times1)$ close-packed \textit{fcc} structure observed by LEED, as shown in Fig.\,1(d). The practically same LEED spots (with slightly increased background) were also observed after intercalation of ~5\,ML of iron underneath graphene layer on Ni(111). Fig.\,1(e) shows the possible crystallographic structure of the graphene/Fe(111)/Ni(111) system obtained after intercalation.

The process of Fe intercalation underneath graphene layer was simultaneously monitored by XPS of principal core levels as well as UPS of the valence band. The present studies of intercalation were performed for 2\,ML and 5\,ML-thick Fe layers on top of the graphene/Ni(111) system. In both cases we found that iron is completely intercalated underneath graphene layer. In the following we will focus on the intercalation process of the 2\,ML-thick Fe film. Fig.\,2 shows XPS spectra of different steps of the preparation of the Fe-based intercalation-like system: (a) Ni $2p_{3/2,1/2}$, (b) C $1s$, and (c) Fe $2p_{3/2,1/2}$. In Fig.\,2(a) all Ni $2p$ emission spectra consist of a spin-orbit doublet ($2p_{3/2,1/2}$) and a well-known satellite structure. The main line is ascribed to a completely screened final state $(c^{-1}3d^{10}4s^{1})$ and the satellite to a two-hole bound state $(c^{-1}3d^{9}4s^{2})$, where $c^{-1}$ stands for the [Ar] core with a $2p$ hole. For the pure Ni(111) film [spectrum 1 in Fig.\,2(a)] the satellite appears with respect to the main lines at 6\,eV higher binding energy, whereas this shift is increased approximately by 0.9\,eV for the graphene/Ni(111) system [spectrum 2 in Fig.\,2(a)]. This effect reflects the altered chemical environment at the interface and is not subject of the present discussion. For the graphene/Ni(111) the C $1s$ XPS spectrum [open circles in Fig.\,2(b)] shows single emission line demonstrating only one carbon phase at the surface (graphene). 

Deposition of the 2\,ML-thick Fe film on top of the graphene/Ni(111) system leads to the reduction of intensities of the Ni $2p_{3/2,1/2}$ [spectrum 3 in Fig.\,2(a)] and the C $1s$ [solid line in Fig.\,2(b)] emission lines. The XPS Fe $2p_{3/2,1/2}$ spectrum in this case has a typical shape characteristic for the metallic iron [dashed line in Fig.\,2(c)].

Annealing of the 2\,ML\,Fe/graphene/Ni(111) system at 600\,K leads to the successful intercalation of the Fe layer underneath graphene. It is supported by the respective changes in the monitored XPS spectra: intensity of Ni $2p_{3/2,1/2}$ spectra practically does not change [spectrum 4 in Fig.\,2(a)], intensity of C $1s$ spectra is restored and coincide with the one for the graphene/Ni(111) spectra [open circles and dash-dot line in Fig.\,2(b)], and intensity Fe $2p_{3/2,1/2}$ emission line reduces in accordance with assumption that graphene layer stays on top of the system and mean free path for electrons emitted from Fe $2p$ level has to be taken into account. However, these changes of intensities of the emission lines can not be simply considered as a proof of a successful Fe intercalation underneath graphene layer, because the practically the same changes in the intensities of XPS lines can be ascribed to the formation of high iron islands on top of graphene layer. In order to rule out this assumption the test experiments on investigation of inert properties of graphene layer were performed as suggested in~\cite{Dedkov:2008a}. Two systems under study were exposed to oxygen at partial pressure $p_{O2}=1\times10^{-6}$\,mbar for 20\,min: 2\,ML of Fe on top of graphene/Ni(111) and system obtained after intercalation of iron atoms in the space between graphene layer and Ni(111). The results are shown in Fig.\,2(c) where Fe $2p_{3/2,1/2}$ XPS spectra for these two cases are shown [compare spectra shown by open squares and open circles in Fig.\,2(c)]. 2\,ML-thick Fe film on top of graphene/Ni(111) is completely oxidized after such oxygen treatment and spectrum is similar to one for magnetite~\cite{Gota:1999}. Opposite to that case the oxygen exposure of the system obtained by intercalation of Fe underneath of graphene layer does not change the Fe $2p_{3/2,1/2}$ XPS spectra. The intensity of the O $1s$ photoemission signal of the graphene/2\,ML\,Fe(111)/Ni(111) system after oxygen exposure is very weak as compared to the one of the 2\,ML\,Fe/graphene/Ni(111) system upon the same treatment [compare spectra 1 and 2 in Fig.\,2(d)]. These facts together with the observation of well ordered LEED spots confirm the formation of graphene-protected intercalation-like system: graphene/2\,ML\,Fe(111)/Ni(111). 

The same scenario is applicable to the description of the UPS spectra of the valence band of the system under study. These results are presented in Fig.\,3 where valence-band photoemission spectra are shown for pure Ni(111) (spectrum 1), graphene/Ni(111) (spectrum 2), 2\,ML\,Fe/graphene/Ni(111) (spectrum 3) and graphene/2\,ML\,Fe(111)/Ni(111) (spectrum 4), and systems after exposure the last ones to the oxygen at the same conditions as mentioned above (spectra 5 and 6). The spectra were obtained with He\,II$\alpha$ radiation in normal emission geometry and are in good agreement with previous data~\cite{Dedkov:2001,Dedkov:2008a,Dedkov:2008b}. From a comparison of the photoemission spectra of graphene/Ni(111) and pure graphite~\cite{Dedkov:2008a} one may conclude that the difference in binding energy of the $\pi$-states amounts to about 2.3\,eV which is in good agreement with the theoretical prediction of 2.35\,eV~\cite{Bertoni:2004}. This shift reflects the effect of hybridization of the graphene $\pi$ bands with the Ni $3d$ bands and, secondary, with the Ni $4s$ and $4p$ states. These results indicate high quality of the graphene monolayer on top of the Ni(111) surface. Deposition of the 2\,ML-thick Fe film leads to the decreasing of the intensity of $\pi$ states of graphene and modification of the emission in the valence band region close to the Fermi level where Ni and Fe $3d$ overlap with each other (compare spectra 2 and 3 in Fig.\,3). Annealing of this system leads to the complete restoring of intensity of the graphene-derived $\pi$ states and new modification of the photoelectron emission in the region of $3d$ emission (compare spectra 2, 3, and 4 in Fig.\,3). Inert properties of the graphene/2\,ML\,Fe(111)/Ni(111) system, as described in the previous paragraph, were tested by exposure this system to oxygen at the same conditions. As one can see there is almost no changes in the photoemission spectra of the valence band region of this system after oxygen adsorption, whereas the UPS spectrum of the system obtained by oxygen deposition on 2\,ML\,Fe/graphene/Ni(111) shows strong contribution of the O $2p$ emission and drastic changes in the valence-band region close to the Fermi level. 

\textit{In conclusion}, we studied the intercalation of thin Fe layers with 2 and 5\,ML thickness underneath graphene layer and formation of the graphene-protected Fe films on Ni(111). The process of intercalation was monitored via XPS of corresponding core levels and UPS of the graphene-derived $\pi$ states in the valence band. Exposure of the formed graphene/2\,ML\,Fe(111)/Ni(111) to large amounts of oxygen does not affect the XPS as well as UPS spectra of the system confirming a successful Fe intercalation and inert properties of the Fe-based intercalation-like system. We suggest that such inert systems with out-of-plane magnetic anisotropy may be used as chemically stable elements in the future spintronics devices. Moreover, \textit{fcc}-Fe layers underneath graphene layer are of special interest for the studies of electronic and magnetic properties of graphene in perpendicular to plane local magnetic field.

This work was funded by the Deutsche Forschungsgemeinschaft, SFB 463, Project B4 and SFB 767, Project C5.

\newpage

\textbf{Figure captions:}
\newline
\newline
\textbf{Fig.\,1.}\,\,(Color online) LEED images of (a) W(110), (b) Ni(111)/W(110), (c) graphene/Ni(111)/W(110) systems, and (d) the system obtained after intercalation of 2\,ML of Fe underneath graphene layer on Ni(111). All images are collected at primary electron energy of 99\,eV. (e) Schematics of a possible crystallographic structure of the graphene/2\,ML\,Fe(111)/Ni(111) system. The corresponding distances between Ni (Fe) atoms in (111) plane and lattice constant of the graphene layer are indicated on the left-hand side of the figure.
\newline
\newline
\textbf{Fig.\,2.}\,\,(Color online) XPS monitoring of Fe intercalation underneath graphene on Ni(111) and inert properties of graphene-based intercalation-like system: (a) Ni $2p_{3/2,1/2}$, (b) C $1s$, (c) Fe $2p_{3/2,1/2}$, and (d) O $1s$ XPS emission lines of systems under study (see text). Spectra are shifted with respect to each other for clarity.
\newline
\newline
\textbf{Fig.\,3.}\,\,(Color online) Valence band UPS spectra of intercalation of Fe underneath graphene on Ni(111) and inert properties of graphene-based intercalation-like system. Spectra are shifted with respect to each other for clarity.

\end{document}